# Understanding the functional impact of copy number alterations in breast cancer using a network modeling approach

Sriganesh Srihari[1,#], Murugan Kalimutho[2,#], Samir Lal[3], Jitin Singla[4], Dhaval Patel[4], Peter T. Simpson[3,5], Kum Kum Khanna[2] and Mark A. Ragan[1,*]

[1] Institute for Molecular Bioscience, The University of Queensland, St. Lucia, QLD 4072, Australia
[2] QIMR-Berghofer Medical Research Institute, Brisbane, QLD 4006, Australia
[3] The University of Queensland, UQ Centre for Clinical Research, Brisbane, QLD 4029, Australia
[4] Indian Institute of Technology Roorkee, Roorkee, Uttarakhand 247667, India
[5] The University of Queensland, School of Medicine, Brisbane, QLD 4006, Australia

[#]joint first authors
*correspondence should be addressed to m.ragan@uq.edu.au

## ABSTRACT

Copy number alterations (CNAs) are thought to account for 85% of the variation in gene expression observed among breast tumours. The expression of *cis*-associated genes is impacted by CNAs occurring at proximal loci of these genes, whereas the expression of *trans*-associated genes is impacted by CNAs occurring at distal loci. While a majority of these CNA-driven genes responsible for breast tumourigenesis are *cis*-associated, *trans*-associated genes are thought to further abet the development of cancer and influence disease outcomes in patients. Here we present a *network-based* approach that integrates copy-number and expression profiles to identify putative *cis*- and *trans*-associated genes in breast cancer pathogenesis. We validate these *cis*- and *trans*-associated genes by employing them to subtype a large cohort of breast tumours obtained from the METABRIC consortium, and demonstrate that these genes accurately reconstruct the ten subtypes of breast cancer. We observe that individual breast cancer subtypes are driven by distinct sets of *cis*- and *trans*-associated genes. Among the *cis*-associated genes, we recover several known drivers of breast cancer (*e.g. CCND1*, *ERRB2*, *MDM2* and *ZNF703*) and some novel putative drivers (*e.g. BRF2* and *SF3B3*). siRNA-mediated knockdown of *BRF2* across a panel of breast cancer cell lines showed significant reduction in cell viability for ER-/HER2+ (MDA-MB-453) cells, but not in normal (MCF10A) cells thereby indicating that *BRF2* could be a viable therapeutic target for ER-/HER2+ cancers. Among the *trans*-associated genes, we identify modules of immune-response (*CD2*, *CD19*, *CD38* and *CD79B*), mitotic/cell-cycle kinases (*e.g. AURKB*, *MELK*, *PLK1* and *TTK*), and

DNA-damage response genes (*e.g. RFC4* and *FEN1*). siRNA-mediated knockdown of *RFC4* significantly reduced cell proliferation in estrogen receptor-negative normal breast and cancer lines, thereby indicating that *RFC4* is essential for cancer cell survival but could also be a useful biomarker for aggressive (ER-negative) breast tumours. Availability: http://bioinformatics.org.au/tools-data/ under NetStrat.

# 1 INTRODUCTION

With an estimated 1.38 million new cases diagnosed in a single year, breast cancer is the most common malignancy affecting women worldwide [1]. Sporadic breast cancer, which accounts for most (93-95%) breast cancer cases, is highly heterogeneous and is now understood to be a collection of diseases with distinct clinical and molecular phenotypes. Gene-expression profiling studies [2,3] have helped delineate this heterogeneity to a certain extent, by identifying at least four major molecular subtypes of sporadic breast cancer (PAM50 subtypes) – basal-like, HER2+, luminal-A and luminal-B. Each of these subtypes is associated with distinct clinical outcomes including patient survival, rates of disease progression and therapeutic response (reviewed in [4]).

Although triggered by mutations in oncogenes such as *PIK3CA* and tumour-suppressor genes such as *PTEN* and *TP53*, breast tumours are largely characterized by amplifications, deletions or rearrangements of chromosomal regions (*i.e.* copy-number aberrations (CNAs)) rather than mutations (see [5]). These somatically acquired CNAs at gene loci account for roughly 85% of the variation in expression (covering ~40% of genes) seen in breast tumours [6]. When these CNAs affect driver genes (oncogenes and tumour suppressors) they further abet the development of cancer and influence disease outcomes in patients. For example, CNAs in *FGFR1* and *ZNF703* arising from aberrations in the chromosomal region 8p12 have been associated with worse prognosis in breast cancer patients [7,8]. Therefore, factoring CNA profiles of genes into the analysis of breast tumours is important to predict disease outcomes accurately and delineate the heterogeneity of breast cancer.

Recent efforts into comprehensive profiling of breast tumours such as by The Cancer Genome Atlas (TCGA) [9] have enabled data integration across diverse datasets (*e.g.* gene expression and CNA). A landmark study by Curtis *et al.* [6] integrating copy number and gene expression in a discovery and validation set of 997 and 995 breast tumours, respectively, identified up to ten subtypes of breast cancer. That study analysed the impact of CNAs on expression variation by grouping genes as *cis*-or *trans*-associated to explain expression variation in these tumours. The expression of *cis*-associated genes is impacted by CNAs at proximal loci, whereas that of *trans*-associated genes is impacted by CNAs at other (distal) sites in the genome. By selecting the top 1000 *cis*-associated genes, constituting several cancer drivers, as features for input to an integrative clustering framework called iCluster [10], ten subgroups of breast cancer (*integrative clusters* 1-10) were identified that showed distinct disease-specific survival for patients. These ten subgroups were subsequently reproduced and validated in an independent dataset of expression profiles from about 7500 tumours [11].

Several methodologies capable of integrating genome-wide CNA and expression profiles have since been developed (reviewed in [12]). For example, Adler *et al.* [13] proposed a stepwise approach, Stepwise Linkage Analysis of Microarray Signatures (SLAMS), in which expression signatures treated as 'phenotype' were associated with CNA signatures treated as 'genotype' to identify potential regulator genes. Ding *et al.* [14] present a visualization system called interactive Genomics Patient Stratification explorer (iGPSe) to visually explore disease subtypes in heterogeneous genomic data. Hofree *et al.* [15] present a framework using mutation profiles to subtype tumours; but being restricted to mutations, the method is less applicable to cancers of the breast and ovary that are largely associated with CNAs.

To understand the role of CNAs in driving breast cancer tumorigenesis, it is important to consider both their *cis*- as well as *trans*-association in impacting the expression landscape of genes. Most identified CNA-drivers in cancers, for example *ERRB2*/HER2+, are *cis*-associated; however, much less is known about *trans*-acting CNA events. Curtis *et al.* [6] note that poor prognosis observed in breast tumours in their integrative cluster 10 is due to overexpression of mitotic genes impacted in *trans* by specific deletion events at distal sites (on chromosome 5q). Consequently, expression changes in these *trans*-associated genes constitute the footprints of *trans*-acting CNAs that influence clinical outcomes.

Here, we seek to identify *cis*- and *trans*-acting genes that drive breast tumours, using a *network*-based approach. We analyse these genes in the context of the ten iCluster breast cancer subtypes as identified by Curtis *et al.* [6], and functionally validate selective genes whose overexpression correlates with poor prognosis in certain breast cancer subtypes. *cis*-Associated genes are typically identified by correlating CNA and expression profiles of genes; however, this typically leads to a long list of genes, many of which are co-expressed with one another. To prioritise our *cis*-associated set we employ a *minimal vertex cover* approach over a network (**Figure 1a**; discussed in detail in Methods). Specifically, we construct the network using correlation between CNA and expression profiles of genes as *node weights* and the co-expression between genes as *interaction weights*, and compute a minimal weighted vertex cover to compose our *cis*-associated genes. To identify *trans*-associated genes, we employ a *network propagation* approach which iteratively computes the influence of CNAs on neighbouring genes in the network (**Figure 1b**; see **Section 4.1** Methods). To validate our identified genes, we reconstruct the known ten clusters of breast cancer as identified by Curtis *et al.* [6]. We follow up on a *cis*-associated gene *BRF2* and a *trans*-associated gene *RFC4* using siRNA-mediated knockdown across a panel of breast cancer cell lines. *BRF2* knockdown reduces cell proliferation by up to 75% in HER2+ cell lines, thereby strongly suggesting that *BRF2* could be an oncogenic driver for these cells. On the other hand, *trans*-associated *RFC4* gene knockdown reduces cell proliferation (up to 60%) in ER-negative normal breast and cancer cell lines, thereby indicating that *RFC4* could be an essential gene, but also a useful biomarker for aggressive (ER-negative) breast tumours.

## 2 RESULTS

### 2.1 Application to subtyping of breast tumours

Selecting only genes common between the PPI and METABRIC discovery datasets produced a network $H_w$ with 118646 interactions among 13038 genes, with 4150 interactions of absolute weights $\geq \omega = 0.40$. Identifying *cis*- and *trans*-associated genes using this network and setting the number of iterations $I = 3$ resulted in $|V_{cis}| = 917$ and $|V_{trans}| = 663$ genes (overlap: $|V_{cis}| \cap |V_{trans}| = 53$), which were used to cluster the tumours.

The *cis*-associated genes produced better clustering (assessed using Silhouette index; Supplementary files) than did the *trans*-associated genes, whereas combining the two sets performed better than either individually. Likewise using $k = 10$ produced better clustering, whereas $k = 8$ merged multiple clusters and $k > 10$ artificially broke apart clusters. **Figures 2 and S2** (clearer figures are available at the Supplemental website) shows disease-specific survival proportions for clusters produced using *cis*-, *trans*- and combined *cis*- and *trans*-associated genes for $k = \{8, 9, 10\}$. The log-rank test *p*-values were highly significant ($p < 0.0001$) in each of these cases, but the finer clusters produced at $k = 10$ (**Figure S2i**) still showed distinct survival patterns thereby representing different clinical subgroups. Therefore, we used $k = 10$ in all our further analyses.

Interestingly the *trans*-associated genes could separate out clusters with poor prognosis (clusters 5 and 7 in **Figure 2**) from the remaining clusters even though this set overlapped with the *cis*-associated genes only by about 8% (=53/663). Upon analyzing the contributions accumulated by these *trans*-associated genes in the poor-prognosis clusters, we found enrichment for cell-cycle and genome-stability functions (18.4% of genes, $p = 2.1 \times 10^{-35}$) with significantly higher contributions for poor-prognosis clusters compared to the remaining clusters (ANOVA test between the means $p < 0.0001$; Bartlett's test: difference between *stddev* $p < 0.0001$) (**Figure 3**), indicating that these genes are associated in *trans* with the poor survival outcomes of these clusters. Curtis *et al.* note that their integrative cluster 10 is *trans*-influenced by several cell-cycle and genome-stability genes that are associated in *trans* with a deletion event at chromosome 5q.

The ten clusters (**Figure 2**) produced from the *cis*- and *trans*-associated genes overlapped with the integrative clusters of Curtis *et al.* with an adjusted Rand index of 0.737 (ARI; a standard method for comparing two sets of clusters; [22]). We computed the overlap (Jaccard index) $J(O_i, IC_j) = |O_i \cap IC_j|/|O_i \cup IC_j|$ between each pair $\{O_i, IC_j\}$ of our cluster $O_i$ and integrative cluster $IC_j$. For $J \geq 0.50$ and $J \geq 0.67$, eight and six of the ten integrative clusters were recovered by our clusters, respectively. Clusters 4 and 8 which showed the worst prognosis at five years (**Table 1**) matched the integrative clusters 10 and 5, respectively. Clusters 1, 6 and 9 which showed a significant drop in survival by ten years (**Table 1**) covered the integrative clusters 1 and 2. These findings agreed with the most-prominent grade, stage and number of positive lymph nodes at diagnosis for the ten clusters (**Table 2; Figure S3**): the aggressive clusters 4, 6 and 8 represented grade 3 diseases with 2 to 4 positive lymph nodes on average.

Regenerating these ten clusters using iCluster [10,11], but using our sets of *cis*- and *trans*-associated genes, matched the Curtis *et al.* clusters with an ARI = 0.69. In particular, clusters 7 and 8 were heterogeneous and produced similar (indistinguishable) survival patterns (**Figure 4a**). But cluster 2, which matches our cluster 6, was more homogeneous and showed a significant drop in survival closer to 10 years. We also generated clusters from the METABRIC Validation dataset (**Figure 4b**), and these matched the Curtis *et al.* Validation clusters with an ARI = 0.49 (amounting to 8 and 6 integrative clusters recovered for $J \geq 0.50$ and $J \geq 0.67$, respectively).

We next compared our clusters with the PAM50 subtypes which are annotated within the TCGA dataset. We obtained 1263 *cis*- and *trans*-associated genes which identified $k = 5$ clusters (**Figure 4c**) with distinct overall survival patterns, and matching PAM50 subtypes with an ARI = 0.67. Cluster 1 was the most homogeneous and matched the basal-like subtype (~87%), whereas cluster 2 matched Her2+ (~76%), cluster 3 was composed of basal-like (23%) and luminal-B (~70%), cluster 4 was composed of luminal-B (~67%) and luminal-A (21%) and cluster 5 the luminal-A (~83%) subtype.

Going back to **Figure 2**, clusters 4 and 8 matched the PAM50 subtypes basal-like and Her2+, whereas clusters 6 and 9 matched luminal-B. This again agrees with Tables 1 and 2: clusters 4 and 8 represented grade 3 diseases and showed the steepest drop in survival cases within five years, which is characteristic of basal-like and Her2+ tumours [23]). Similarly, clusters 6 and 9 also represented grade 3 disease but showed a significant drop in survival cases by ten years, which has been observed for luminal-B tumours (distant relapse) [23].

### 2.2 In-depth analysis of *cis*- and *trans*-associated genes in the ten clusters

Among the *cis*-associated genes we recovered were many known breast cancer drivers including *CCND1*, *ERBB2*, *MYC*, *MDM2* and *ZNF703* [6,8,24] which showed considerable gains (**Figures S8, S9**) with their over-expression significantly associated with poor survival. For example, *ZNF703* is a regulator of cell adhesion, migration and proliferation. Over-expression of *ZNF703*, attributed to amplification of the 8p12 region, has been frequently observed in estrogen-receptor positive (luminal) tumours with poor outcomes [8]. Accordingly, we observed amplification and over-expression of *ZNF703* in clusters 1 and 3 (**Figure S8**) (matching integrative clusters 1 [luminal-B], 8 [luminal-A] and 9 [mixed]), which showed poor survival at ten years' follow-up (**Table 1**). Over-expression of *ZNF703* ($\geq 75$ percentile) was noted in 94% of these tumours (**Figure S4**).

The three splicing factors *SF3B3*, *SF3B4* and *SF3B5* showed contrasting patterns of CNAs (**Figures S8, S9**), with *SF3B3* and *SF3B5* showing losses whereas *SF3B4* showing considerable gains across the tumour clusters. However, only the amplifications and corresponding over-expression of *SF3B3*, which was confined particularly to cluster 4 (basal-like), associated with poor prognosis (**Figures 5a**). In particular, 40% of tumours showing very high expression ($\geq 85$ percentile) for *SF3B3* and poor prognosis originated from cluster 4, which is predominantly basal-like (**Figure 5b** – in particular, compare the survival curves for very high (red) and low (green) expression levels; **Figure S5**). A recent study [25] has associated the overexpression of *SF3B3* with poor prognosis in luminal (estrogen receptor-positive) tumours, and our results complement this study by associating *SF3B3* with also the poor prognosis of (a subset of) basal-like (estrogen receptor-negative) tumours.

Another potentially interesting gene is *BRF2*, which is ranked second (behind only *ERBB2*) in our list of *cis*-associated genes in breast cancer. *BRF2* is a subunit of the RNA polymerase II complex, which is required for transcription. Two earlier reports [37, 38] linked aberrant expression of *BRF2* with deregulation of transcription in cancer cells, and therefore suggest *BRF2* as a putative oncogene. In particular, a recent report [39] demonstrates that overexpression of *BRF2* is required for the development of lung squamous cell carcinoma, and its activation is observed in >35% of lung cancers. We did not observe a significant association between the expression of *BRF2* and survival in breast cancers. Nevertheless, Western blot analysis showed higher protein expression for BRF2 in several breast cancer cell lines, in particular HER2+ cell lines (**Figure 6a**). siRNA-mediated knockdown of *BRF2* showed reduction in cell proliferation in some of these cell lines, in particular up to 75% reduction in the ER-/HER2+ cell line MDA-MB-453 (**Figures 6b & 6c**). But, the effect was not observed in normal MCF10A cells, suggesting that BRF2 might be a viable therapeutic target. Recent reports [40,41] support this view by suggesting that *BRF2* could act as an 'alternative driver' for HER2+ breast cancers that do not express the HER2/*ERBB2* gene.

The analysis of *trans*-associated genes revealed enrichment for primarily two kinds of cellular processes (**Figures S6 and S7**): (i) immune-response functions including regulation of T-cell differentiation and activation (5.1%, $p = 6.0 \times 10^{-17}$), and antigen presentation (1.8%, $p = 1.5 \times 10^{-8}$); and (ii) mitotic functions including M-phase (9.3%, $p = 8.1 \times 10^{-22}$), G2/M DNA-damage checkpoint (2.4%, $p = 1.2 \times 10^{-5}$), spindle checkpoint (0.8%, $p = 1.5 \times 10^{-3}$) and apoptosis (12.3%, $p = 3.9 \times 10^{-12}$). These genes showed distinct contribution values between the clusters (ANOVA: difference between means $p < 0.0001$; Bartlett's test: difference between *stddev* $p < 0.0001$). In particular, the immune-response genes showed high contribution values for cluster 3 (matching integrative cluster 4 (~40%)), suggesting a tumour subgroup *trans*-influenced by these genes.

Among the mitotic genes were cell-cycle kinases *AURKB*, *CDK1*, *CHEK1, MELK*, *PLK* and *TTK* with the master regulator FOXM1 central to the mitotic network. These genes showed the highest contribution values for cluster 4 (matching integrative cluster 10 (~85%) and basal-like tumours), which displayed the worst prognosis at five years follow-up (**Figure 2** and **Table 1**). Several of these are being explored as therapeutic targets in basal-like/triple-negative breast tumours (reviewed in [4,26]). In particular, knockdown of FOXM1 in specific cell lines including MDA-MB-231 reduces tumour-cell colony formation and increases sensitivity to the drug doxorubicin *via* reduction in the transcription of DNA-damage response (DDR) genes [27,28]. More recently, we have shown that inhibition of TTK significantly synergized with docetaxel in the treatment of aggressive triple negative breast cancer xenografts and exhibit as potential target in breast cancer [42]. Since most cell-cycle kinases have already been explored, we chose to follow up on *RFC4*, a replication factor gene and a downstream target of FOXM1. Overexpression of *RFC4* was observed predominantly in clusters 4, 5 and 9 (basal-like and luminal-B) and correlated with poor survival in tumours (**Figure 5a** and **c** – in particular, compare the curves for very high (red) and low (green) expression levels). Western blot analysis showed higher expression for *RFC4* in breast cancer cell lines compared to the near-normal mammary epithelial cell line MCF10A (**Figure 7a**). Each of the two siRNAs against RFC4 reduced the expression of RFC4 by ≥ 50% compared to scrambled siRNA used as control (Figure **7b**). Knockdown of RFC4 showed up to 70% reduction in cell proliferation in cancer cell lines with a higher trend visible for ER-negative (basal-like) cell lines, in particular BT549 and MDA-MB-157 (**Figure 7c**). However, the knockdown suppressed proliferation also of MCF10A cells, up to 50%. These observations show that breast cancer cells exhibit higher expression of RFC4, whereas knockdown of RFC4 is lethal predominantly in ER-negative cell lines, which includes MCF10A. This also agrees with a recent large-scale cell-essentiality (siRNA knockdown) screen across 72 cancer cell lines [29] in which *RFC4* is more essential in the ER-negative lines MDA-MD-231, MDA-MB-157 and BT549 compared to the ER-positive lines MCF7 and T47D. *RFC4* encodes the fourth-largest subunit of human replication factor C, a multimeric protein complex consisting of five subunits that functions as a clamp loader to load PCNA onto DNA during DNA synthesis [30, 31]. RFC4 also functions in DDR and checkpoint control through interaction with the DNA-damage checkpoint complex Rad17, 9-1-1 [31,32]. Therefore, high expression of RFC4 could be reflective of high proliferation rate and higher number of proliferative cells in ER-negative cell lines. Furthermore, several recent reports have associated overexpression of *RFC4* with poor prognosis in colorectal [33] and hepatocellular [34] cancers. Putting our findings together with these literature reports, we suggest that the expression of *RFC4* is useful as a biomarker for aggressive breast tumours with its knockdown causing considerable reduction in proliferation of ER-negative cells. However, because *RFC4* is an essential gene even for normal (MCF10A) cells, engaging it as a therapeutic target could be cytotoxic to some extent.

We found 30 of the 444 cancer genes reported in the COSMIC census [43], and 6 of 37 genes affected by amplifications, deletions or rearrangements from Vogelstein's list of cancer genes [44], to be among our *cis*-associated genes (**Table S3**). Most of these known cancer genes, *e.g. CCND1* and *MDM2*, are affected by copy-number amplifications. We also found that some genes with roles in DDR, including *FANCG*, *MLH1* and *PALB2*, were deleted. FANCG, a Fanconi anaemia group protein involved in the repair of DNA crosslinks, works closely with BRCA2. MLH1 is involved in the repair of DNA mismatches. PALB2 plays a crucial role in recruiting BRCA2 and RAD51 to sites of DNA breaks during homologous-recombination-mediated DSB repair. These results suggest that apart from mutation and epigenetic silencing of these DDR genes, genomic deletions play an important role in the development of breast cancers.

## 3 CONCLUSION

Although breast cancers are triggered by somatic mutations in oncogenes such as *PI3KCA* and tumour-suppressor genes such as *PTEN* and *TP53*, their genomic landscape is dominated by chromosomal amplifications and deletions. These CNAs contribute to the development of breast cancer, and substantially influence survival and drug-response patterns in patients. Here, using a cohort of ~2000 breast tumours obtained from a recent comprehensive study [6], we develop a network-based approach to decipher this impact of CNAs on the expression landscape of genes by identifying *cis*- and *trans*-associated genes in breast cancers. To validate our sets of genes, we employ them to subtype the cohort into ten distinct clusters which closely match that from [6]. We note that individual clusters thus identified are driven by distinct sets of genes, for example by overexpression of *ZNF703* in luminal clusters.

Interestingly *trans*-associated genes, which are influenced by CNAs at distal foci and overlap very little (<8%) with *cis*-associated genes, show significant differences between their involvement in the clusters. This strongly suggests that CNAs in *trans* also play an important role in influencing differences between the clusters. Our network approach identified 71 *trans*-associated genes that were also highlighted in [6]. Among these, we found primarily two distinct functional modules – of mitotic/cell-cycle and immune-response genes (**Figures S6 and S7**). Several of these cell-cycle/mitotic genes include kinases (*e.g.* AURKB, CDK1, MELK, PLK and TTK) which have been extensively explored as therapeutic targets (reviewed in [4,26]), genes involved in DNA replication (*e.g. MCM2, MCM4, CDC7, CDC45* and *RFC4*), and also DDR genes (*e.g. RFC4* and *FEN1*) with the master regulator FOXM1 at their core.

Here, we demonstrate an effective pipeline from modeling to analysis and prediction, and experimental validation of our results. Our modeling framework presents an elegant way to decipher the functional influence patterns of CNAs, and is an effective computational alternative to exhaustive analysis of the genomic architecture and using an eQTL framework as adopted by Curtis *et al.* [6].

## 4 MATERIALS AND METHODS
### 4.1 Identifying *cis*-associated genes

We assemble a protein interaction (PPI) network $H = (V, E)$, where $V$ is the set of genes (proteins) and $E$ is the set of interactions (unweighted and undirected) between these genes. Let $T$ be the set of tumours, and $G$ [$n$ X $m$] and $C$ [$n$ X $m$] represent the gene-expression and CNA profiles spanning $n$ genes from these $|T|=m$ tumours, respectively. Using gene-expression profiles $G$, each interaction $(g_x, g_y) \in E$ is assigned a weight $w(g_x, g_y) \in [-1,1]$ as the co-expression (as Pearson correlation) observed between $g_x$ and $g_y$ across the $m$ tumours, resulting in a weighted network $H_w$.

For each gene $g_x \in V$, we compute the node weight $NV(g_x)$ in $H_w$ as

$$NV(g_x) = 1 - |\rho(G[g_x],C[g_x])|, \qquad (Eq.\ 1)$$

where $0 \leq |\rho(G[g_x],C[g_x])| \leq 1$ is the correlation between CNA and expression profiles of $g_x$ across the $m$ tumours. To identify *cis*-associated genes, we find the set of genes with minimal node weights which *cover* all the interactions in $H_w$, or in other words, we find the *minimum weight vertex cover* (MWVC) of $H_w$, formulated as:

MWVC of $H_w$: Find a subset $V_{cis} \subseteq V$ such that for every interaction $(g_x, g_y) \in E$ at least one of $g_x$ or $g_y$ is in $V_{cis}$ and the total weight $\sum_{g_x \in V_{cis}} NV(g_x)$ is minimized.

In general, MWVC is NP-complete [16], but here we employ a heuristic approach to obtain a suboptimal solution, as follows. We pick an interaction $(g_x, g_y)$ and select $g_x$ into $V_{cis}$ if $NV(g_x) \leq NV(g_y)$, marking all interactions incident on $g_x$ as *covered*. We repeat this until all interactions are covered, to identify the set of *cis*-associated genes $V_{cis}$.

CNAs of *cis*-associated genes are strongly associated with expression variation, and for any *cis*-associated gene $g_x$ the correlation $\rho(G[g_x],C[g_x])$ captures this association. Here, we use the top $\alpha$ genes in $V$ with high correlation as our starting point. If any two of these genes $g_x$ and $g_y$ are also strongly co-expressed, as determined by $|w(g_x, g_y)| \geq \omega$ (a threshold), and if $\rho(G[g_x],C[g_x]) \geq \rho(G[g_y],C[g_y])$ (modelled as $NV(g_x) \leq NV(g_y)$), then we select only $g_x$ into $V_{cis}$. This reduces the *redundancy* of our selection (*i.e.* we select fewer genes) because we always pick gene $g_x$ with higher $\rho(G[g_x],C[g_x])$ when $g_x$ and $g_y$ show similar expression profiles (**Figure 1a**) ($\alpha$ and $\omega$ are empirically determined as explained later).

### 4.2 Identifying *trans*-associated genes

Unlike *cis*-associated genes, the *trans*-associated genes do not show a strong correlation between CNA and expression profiles, but instead their expression is influenced by CNAs of other (*cis*-associated) genes. We compute this influence on *trans*-associated genes using the interactions in $H_w$. However, since each tumour has a distinct CNA profile, the pattern of influence exerted by genes on each other varies considerably between tumours. Therefore, we compute the influence between genes for each tumour individually, and then select genes that accumulate a high overall influence across all tumours, as our set of *trans*-associated genes.

For a tumour $T_s \in T$, each gene $g_x$ in $H_w$ is assigned an initial node weight $NV^{(0)}[T_s](g_x) = C[x,s]$ obtained from the CNA profiles $C$. Using the interactions in $H_w$, we propagate these node weights to neighbouring genes iteratively across the network (**Figure 1b**). In each iteration $i \geq 1$, the contribution $CV^{(i)}[T_s](g_x)$ that $g_x$ receives from its neighbours is computed as the interaction-weighted sum of node weights of neighbours, given by

$$CV^{(i)}[T_s](g_x) = \frac{\sum_{g_z \in N(g_x)} \varphi(w(g_x,g_z)) \cdot NV^{(i-1)}[T_s](g_z) \cdot |w(g_x,g_z)|^i}{|N(g_x)| + \lambda(g_x)}, \quad \text{(Eq. 2)}$$

where $N(g_x)$ is the set of neighbours of $g_x$ in $H(G)$, $\varphi(w(g_x, g_z))$ returns the sign of $w(g_x, g_z)$, that is whether positive or negative, and $\lambda(g_x)$ is added to penalize genes with very few neighbours, given by:

$$\lambda(g_x) = \max\left\{0, \frac{\sum_{g_z \in V} |N(g_z)|}{|V|} - |N(g_x)|\right\}. \quad \text{(Eq. 3)}$$

At each iteration, the above computation is performed for every gene $g_x$ in the network, and at the end of the iteration the node weights are synchronously updated, $NV^{(i)}[T_s](g_x) = NV^{(i-1)}[T_s](g_x) + CV^{(i)}[T_s](g_x)$ for all genes. We perform $i = I$ iterations, but dampen the contributions using an exponential decay of the interaction weights (since $|w(g_x, g_z)| \leq 1$ in Equation 2).

After executing the above procedure for every tumour $T_s \in T$, we sum the contributions for each gene $g_x$ across all tumours,

$$CV^{(I)}[T](g_x) = \sum_{T_s \in T} |CV^{(I)}[T_s](g_x)|, \quad \text{(Eq. 4)}$$

and select genes with $CV^{(I)}[T](g_x) > 0$, which is the union of all genes that accumulate non-zero contributions, as our set of *trans*-associated genes $V_{trans}$.

Each gene $g_x$ receives contributions from its neighbours $g_z \in N(g_x)$ relative to the CNA of $g_z$ and the co-expression between $g_x$ and $g_z$. If the CNA of $g_z$ affects the expression level of $g_z$ and if $g_z$ is co-expressed with $g_x$, then we assume that the CNA of $g_z$ influences the expression of $g_x$, thereby making $g_x$ a *trans*-associated gene. Equation 2 captures this influence as the co-expression-weighted sum of CNAs from all neighbours $g_z$ for which $|w(g_x, g_z)| \geq \omega$. Through iterations, we account for neighbours up to $I$ hops away (see **Figure 1b**). Genes that accumulate non-zero contributions (influence) are selected into $V_{trans}$.

### 4.3 Validation of *cis*- and *trans*-associated genes for breast cancer subtyping

One way to validate our identified *cis*- and *trans*-associated genes and to understand their biological relevance is to use them for subtyping breast cancers. Here we use the expression profiles of the two sets of genes, $V_{cis}$ and $V_{trans}$ as input features to a *k*-means clustering to identify tumour subgroups $O = \{O_1,..., O_k\}$. We test $k \in [2, 15]$ and choose the $k$ giving the best Silhouette index (a measure to assess clustering performance). We then validate these tumour subgroups using Kaplan-Meier survival plots.

### 4.4 Datasets

*4.4.1 METABRIC dataset*

Clinical annotations, gene expression and CNA profiles for the discovery (997) and validation (995) set of tumours from the Curtis *et al.* [6] study were obtained by permission from the METABRIC consortium. Genes altered for copy number were located on the circular binary segmentation (CBS)-smoothed CNA segments from this dataset based on UCSC (h19) gene annotations [17] and processed using R packages [18], to generate a continuous log-ratio (CBS) smoothed CNA matrix for genes, with positive values meaning gains or amplifications and negative values meaning homozygous or heterozygous deletions.

*4.4.2 TCGA dataset*

Clinical annotations, gene expression and CNA profiles for 597 breast tumours were downloaded from the TCGA website (http://cancergenome.nih.gov). Based on molecular intrinsic (PAM50) subtyping these accounted for 81 basal-like, 58 HER2+, 132 luminal B, 235 luminal A, 8 claudin-low tumours, with the remaining 83 unclassified.

*4.4.3 Protein interaction dataset*

*Homo sapiens* PPI data were downloaded from iRefIndex [19] which consolidates PPI data from multiple sources. Only those interactions annotated for direct physical or annotated reactions were selected. Removing self-, duplicate, false-positive interactions [20,21] resulted in a dense PPI network containing 176574 interactions among 19743 proteins (average node degree 17.81).

## 4.5 Setting $\alpha$ and $\omega$

Parameter $\omega$ identifies interactions in $H_w$ that we wish to account for while determining the sets $V_{cis}$ and $V_{trans}$. We cover only those interactions with weights at least $\omega$ for computing $V_{cis}$. Likewise, we use only these interactions to determine the neighbour contributions for computing $V_{trans}$. In accordance with our assumption that the expression of a *trans*-associated gene $g_x$ is influenced by CNAs of its neighbours, reflected as co-expression with its neighbours, we wish to account for only those neighbours $g_y$ whose CNA patterns explain their co-expression with $g_x$. We found that using a cut-off $\omega = 0.40$ on gene co-expression accounted for 4150 interactions $(g_x, g_y)$ of which ~83% (3443) showed CNAs in at least one of $g_x$ or $g_y$, and their expression levels correlated with these CNAs with $r = 0.69$ (**Figure S1**).

As noted in the Curtis *et al.* study, determining the top $\alpha$ *cis*-associated genes is somewhat arbitrary, so here we assessed the top $P \in [5, 25]$ percentile of genes with significant Benjamini-Hochberg *p*-values for CNA-to-expression correlation, and found $P = 10$, accounting for the top $\alpha = 1304$ genes, was the lowest $P$ sufficient to retain the 4150 interactions.

## 4.6 Experimental validation of RFC4 and BRF2 using siRNA-mediated knockdown

Breast cancer cell lines were obtained from ATCC (Manassas, VA, USA) and cultured routinely per ATCC instructions. All cell lines were regularly tested for mycoplasma and authenticated using Short Tandem Repeat (STR) profiling by the QIMR scientific services core facility. For siRNA experiments, 10 nM siRNAs were reverse transfected using Lipofectamine RNAiMAX (Life Technologies, Carlsbad, CA).

We obtained the following siRNAs Shanghai Gene Pharma (Shanghai):

siRFC4_S1: 5′-GACGUACCAUGGAGAAGGAGUCGAA-3′;

siRFC4S2: 5′-GCAGCAGUUAUCUCAGAAUUGUUAA-3′;

siBRF2_S1: 5′-GCAGGGAUGACUAUAGGUAGGAGAG-3′;

siBRF2_S2: 5′-GCAGUUGCCACCAACAUUUGAGGAT-3′;

siBRF3_S3: 5′-CUGCCUCACGGUUGCUGUUGCAGAC-3′ and

siSC: 5′-UUCUCCGAACGUGUCACGUTT-3′.

Six days after siRNA knockdown, cell viability was determined using the CellTiter 96® AQueous One Solution per the manufacturer's instructions (Promega, Fitchburg WI, USA). For immunoblotting, standard protocols were used and membranes were probed with antibodies against RFC4 (anti-RFC4 [N1C3]), BRF2 (anti-BRF2 [N1C1]) (GeneTex CA, USA) and γ-tubulin (Sigma-Aldrich, Sydney), and developed using Chemiluminescence Reagent Plus (Millipore, Billerica, MA, USA).

All data and associated analysis are available at: http://bioinformatics.org.au/tools-data/ under NetStrat.

## ACKNOWLEDGEMENTS

We thank METABRIC (Molecular Taxonomy of Breast Cancer International Consortium) for granting us access to the breast cancer dataset. We thank Dr Alison Anderson for useful discussions.

*Funding*: This work is supported by an Australian National Health & Medical Research Council (NHMRC) grant 1028742 to PTS and MAR. MK is supported by a Cancer Council Queensland (CCQ) Project Grant (1087363), KKK is a NHMRC Senior Principal Research Fellow (ID 613638).

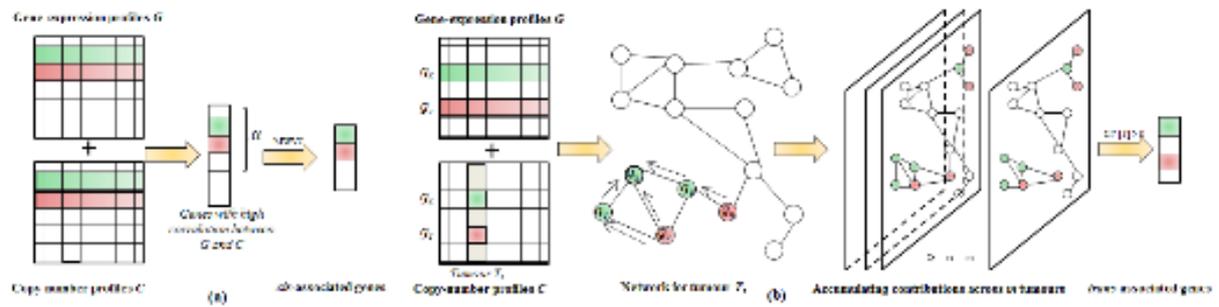

**Figure 1: Identifying *cis*- and *trans*-associated genes. (a)** *cis*-associated genes are identified by correlating expression and CNA profiles. We choose the top-α genes that show high correlation, and identify a vertex cover subset (MWVC) that is representative of these top genes; and **(b)** *trans*-associated genes are identified by propagating node weights across neighbouring genes in the network. Gene $g_x$ receives contributions from the 2-hop neighbour $g_q$ *via* the 1-hop neighbour $g_p$ when the iteration reaches $i = 2$, because $g_p$ would have received contributions from $g_q$ in iteration $i = 1$.

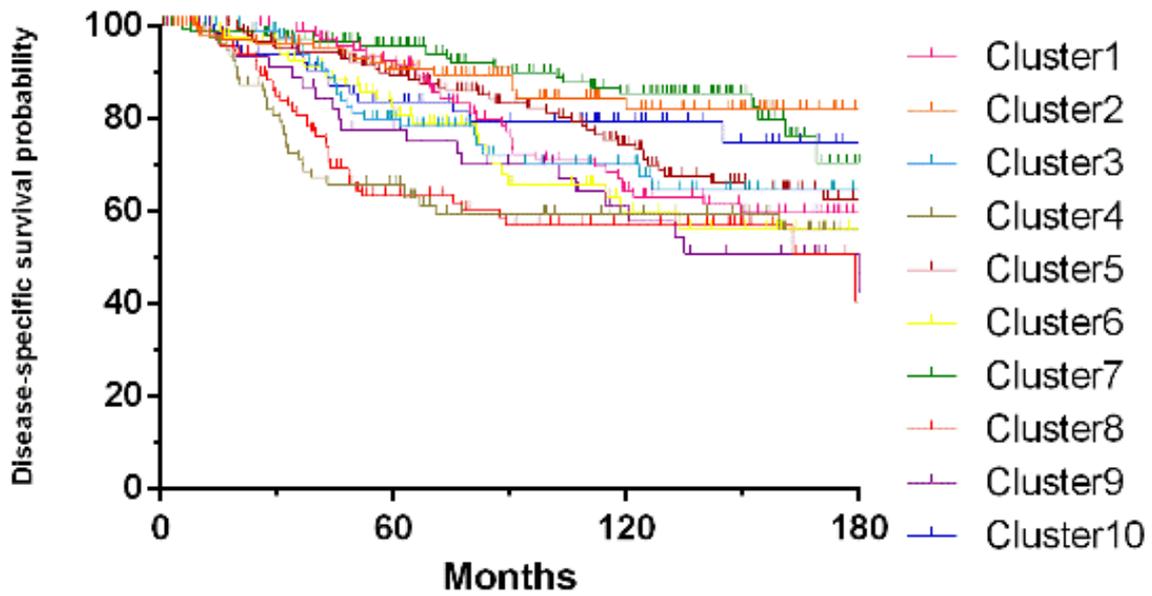

**Figure 2:** Kaplan-Meier plots of disease-specific survival (truncated at 15 years) for clusters identified using *cis*- (917), *trans*- (663) and combined *cis*- and *trans*-associated (1527) genes (arranged horizontally) for $k = 10$ clusters from the Discovery dataset (998 tumours). Log-rank test *p*-value in each of the cases was significant ($p < 0.0001$). A clearer figure is available from Supplementary website. Please also see Figure S2.

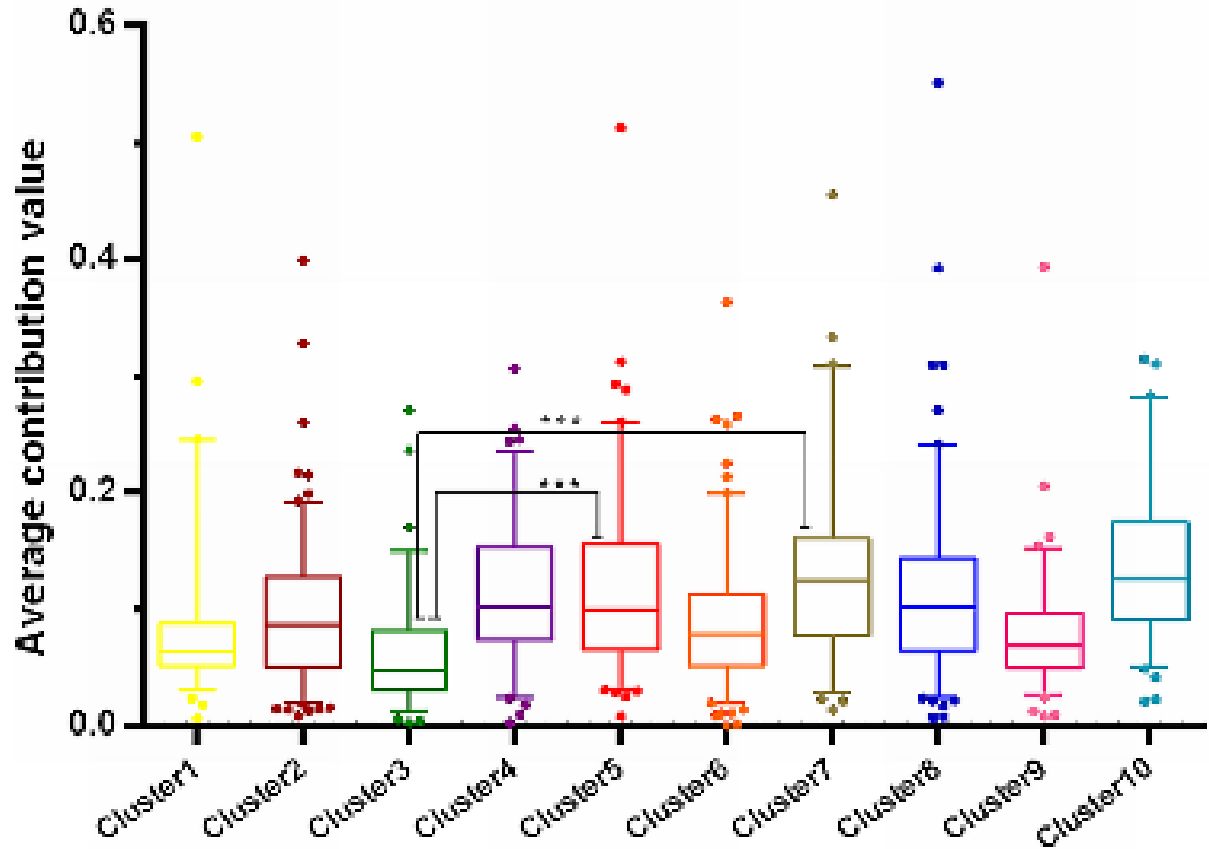

**Figure 3:** Contribution values accumulated by *trans*-associated genes involved in cell cycle and genome stability (genes highlighted also in Curtis *et al.* [6]). The clusters show significant differences between their contribution-means (ANOVA test: *p*-value < 0.0001). Clusters 5, 7 and to some extent 4 and 10 show higher contributions for these genes and the clusters show poorer prognosis, whereas cluster 3 with the lowest contribution shows better prognosis (Figure 2).

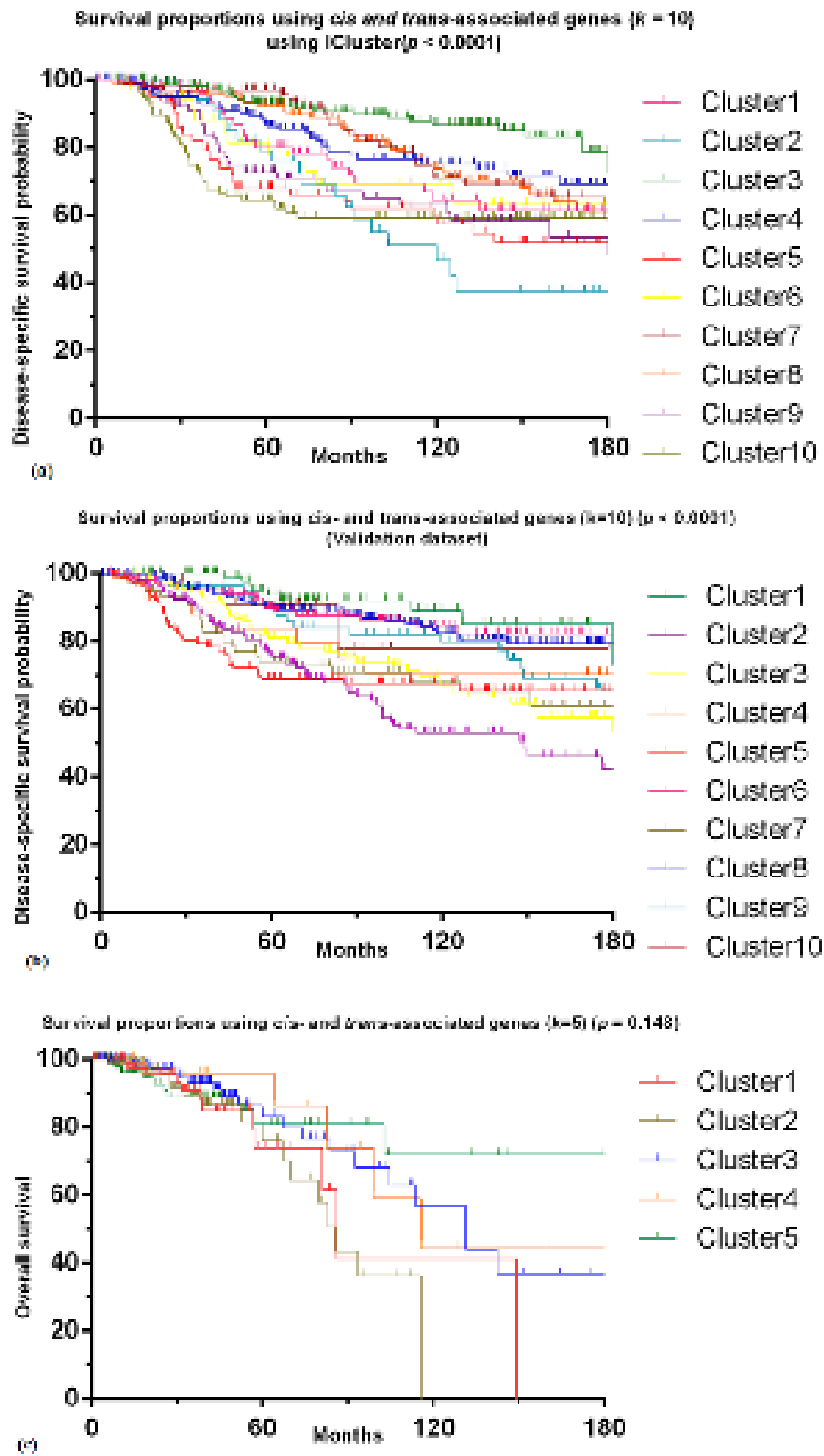

**Figure 4:** Validation on the METABRIC Validation and TCGA cohorts [6]. **(a)** Clusters generated using iCluster [10]. **(b)** Disease-specific survival plots for $k = 10$ generated using the 1527 *cis-* and *trans*-associated genes on the Validation dataset (995 tumours). **(c)** Overall survival plots for $k = 5$ generated using 1263 *cis-* and *trans*-associated genes on the TCGA dataset (597 tumours). Log-rank test $p$-value was not significant ($p = 0.148$), but this could be because of very few deceased cases (about 135) in the TCGA cohort.

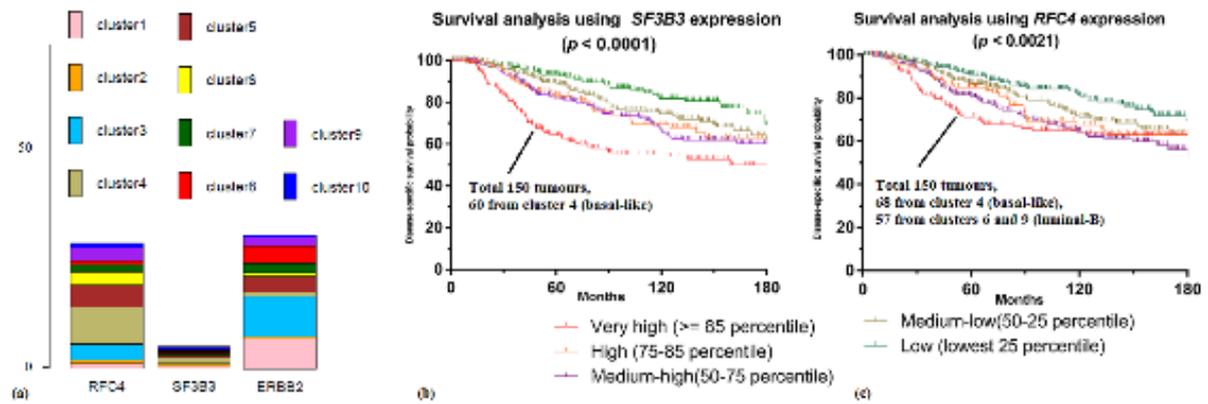

**Figure 5**: **(a)** Fraction (%) of tumours (out of 997) with amplifications for *RFC4* and *SF3B3* in the ten clusters (*ERBB2* is a known oncogene); match cluster numbers with Figure 2. Amplifications for *SF3B3* and *RFC4* were most common for cluster 4. **(b) & (c)** Survival proportions for disease-specific survival in patients stratified by percentiles of expression levels of **(b)** *SF3B3* and **(c)** *RFC4*. Over-expression of these genes correlates with poor survival; ~40% of tumours showing very high expression ($\geq$ 85 percentile) come from cluster 4, which represents predominantly basal-like tumours. Compare red (very high) and green (low expression) curves for better indication of the differences in survival.

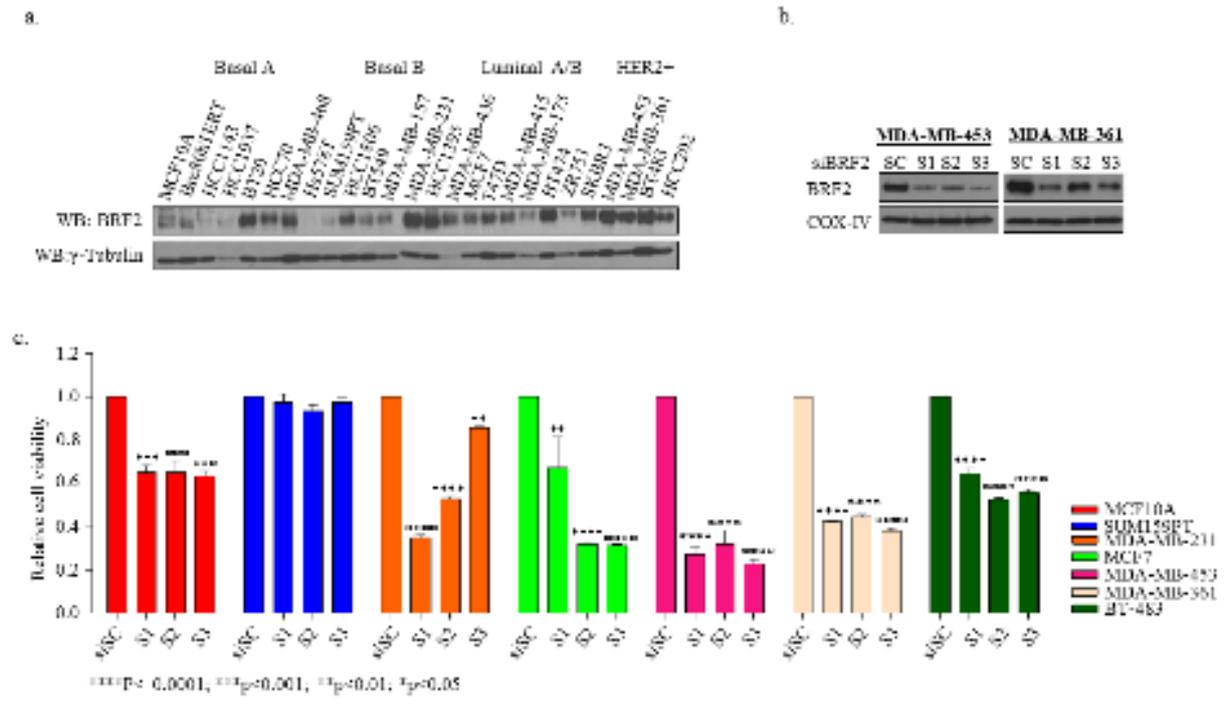

**Figure 6:** Western blot and siRNA-mediated knockdown of *BRF2* in breast cancer cell lines: **(a)** basal expression level of *BRF2* across a panel of cell lines; **(b)** knockdown efficiency using siRNAs S1 and S2 for *BRF2* (SCR: scrambled siRNA used as control); and **(c)** cell viability upon BRF2 knockdown (****$p<0.0001$; ***$p<0.001$; **$p<0.01$).

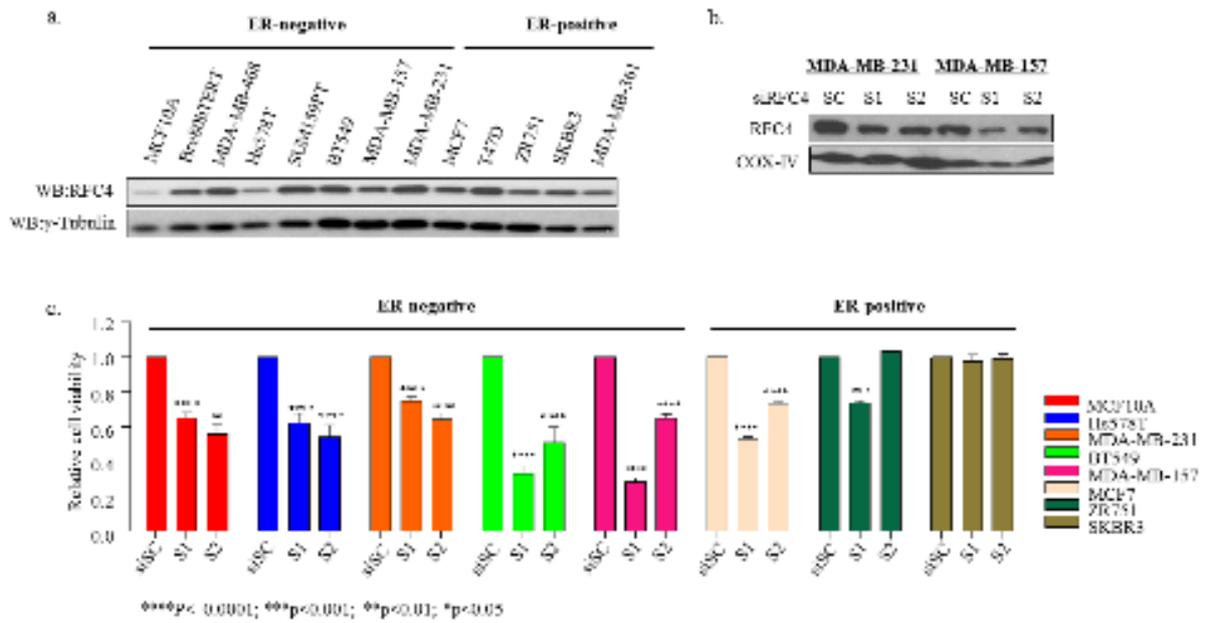

**Figure 7:** Western blot and siRNA-mediated knockdown of *RFC4* in ER-negative and ER-positive breast cancer cell lines: **(a)** basal expression level of *RFC4* across a panel of cell lines; **(b)** knockdown efficiency using siRNAs S1 and S2 for *RFC4* (SCR: scrambled siRNA used as control); and **(c)** cell viability upon RFC4 knockdown (****$p<0.0001$; ***$p<0.001$; **$p<0.01$).

**Tables**

| Cluster | 1 | 2 | 3 | 4 | 5 | 6 | 7 | 8 | 9 | 10 |
|---|---|---|---|---|---|---|---|---|---|---|
| #Risk | 101 | 114 | 92 | 81 | 150 | 89 | 154 | 102 | 47 | 66 |
| #Deaths | 33 | 15 | 23 | 31 | 41 | 24 | 20 | 39 | 20 | 14 |
| #S(60) | 91.3 | 90.7 | 79.7 | 65.8 | 89.2 | 80.7 | 95.7 | 63.2 | 77.5 | 83.5 |
| #S(120) | 64.4 | 81.9 | 70.1 | 59.4 | 74.3 | 59.9 | 85.3 | 56.8 | 57.8 | 79.1 |

**Table 1:** Number of patients at risk and total deaths for the ten clusters of Figure 2. $S(X)$ is the % of patients surviving at $X$ months.

| Cluster | 1 | 2 | 3 | 4 | 5 | 6 | 7 | 8 | 9 | 10 |
|---|---|---|---|---|---|---|---|---|---|---|
| Grade | 2 | 2 | 3 | 3 | 2 | 3 | 2 | 3 | 3 | 2 |
| Stage | 0 | 2 | 3 | 3 | 2 | 2 | 2 | 2 | 1 | 2 |
| #Lymph | 1 | 2 | 3 | 2 | 2 | 4 | 2 | 4 | 2 | 2 |

**Table 2:** Most-prominent grade, stage and number of positive lymph nodes at diagnosis for patients in the ten clusters of Figure 2 (also see **Figure S3**).